\documentclass[aps,twocolumn,groupedaddress,amsmath,amssymb]{revtex4}
\usepackage[pdftex]{graphicx}
\usepackage{amssymb}
\usepackage{bm}
\begin{document}
\setcounter{page}{1}
\title[]{Extended Klein-Gordon Action, Gravity and Non-Relativistic Fluid}
\author{Mokhtar Hassa\"{\i}ne}\email{hassaine-at-inst-mat.utalca.cl}
\affiliation{Instituto de Matem\'atica y F\'{\i}sica, Universidad de
Talca, Casilla 747, Talca, Chile,}
\affiliation{Centro~de~Estudios~Cient\'{\i}ficos~(CECS),
~Casilla~1469,~Valdivia,~Chile.}

\begin{abstract}
We consider a scalar field action for which the Lagrangian density
is a power of the massless Klein-Gordon Lagrangian. The coupling of
gravity to this matter action is considered. In this case, we show
the existence of nontrivial scalar field configurations with
vanishing energy-momentum tensor on any static, spherically
symmetric vacuum solutions of the Einstein equations. These
configurations in spite of being coupled to gravity do not affect
the curvature of spacetime. The properties of this particular matter
action are also analyzed. For a particular value of the exponent,
the extended Klein-Gordon action is shown to exhibit a conformal
invariance without requiring the introduction of a nonminimal
coupling. We also establish a correspondence between this action and
a non-relativistic isentropic fluid in one fewer dimension. This
fluid can be identified with the (generalized) Chaplygin gas for a
particular value of the power. It is also shown that the
non-relativistic fluid admits, apart from the Galileo symmetry, an
additional symmetry whose action is a rescaling of the time.

\end{abstract}

\maketitle

\section{Introduction}
The fundamental tenet of General Relativity is the manifestation of
the curvature of spacetime produced by the presence of matter. This
phenomena is encoded through the Einstein equations that relate the
Einstein tensor (with or without a cosmological constant) to the
energy-momentum tensor of the matter,
\begin{eqnarray}
G_{\mu\nu}+\Lambda g_{\mu\nu}=\kappa\,T_{\mu\nu}.
\label{Einsteineqs}
\end{eqnarray}
Since the energy-momentum tensor depends explicitly on the metric,
both sides of the equations must be solved simultaneously.

In the first part of this paper, we address the following question:
for a fixed geometry solving the vacuum Einstein equations, is it
possible to find a matter source coupled to this spacetime without
affecting it. Concretely, this problem consists of examining a
particular solution of the Einstein equations (\ref{Einsteineqs})
for which both sides of the equations vanish, i. e.
\begin{eqnarray}
G_{\mu\nu}+\Lambda g_{\mu\nu}=0=\kappa\,T_{\mu\nu}.
\label{StealthEinsteineqs}
\end{eqnarray}
In three dimensions, such gravitationally undetectable solutions
have been obtained in the context of scalar fields nonminimally
coupled to gravity with a negative cosmological constant \cite{S3d}.
Recently, the same problem has been considered in higher dimensions
but with a flat geometry \cite{Ayon-Beato:2005tu}.

In our case, we show that for any static, spherically symmetric
vacuum solutions of the Einstein equations (eventually with a
cosmological constant), a particular matter source action can be
coupled to this geometry yielding to non-trivial solutions of the
equations (\ref{StealthEinsteineqs}). This matter source is
described in $(D,1)$ dimensions by an extended Klein-Gordon action,
\begin{eqnarray}
S_{\alpha}=\int
d^D\vec{x}\,dt\,\sqrt{-g}\left(g^{\mu\nu}\,\nabla_{\mu}
\theta\,\nabla_{\nu}\theta\right)^{\alpha}, \label{action}
\end{eqnarray}
where $\theta$ is the dynamical field and $\alpha$ is a real
parameter whose range will be fixed later. For $\alpha=1$, this
action reduces to the well-known massless Klein-Gordon action.

The derivation of solutions of the equations
(\ref{StealthEinsteineqs}) is shown to be equivalent of finding a
null vector field $k^{\mu}$ (for the vacuum metric $g_{\mu\nu}$)
which derives from the scalar field $\theta$. For a static,
spherically symmetric spacetime, the derivation of vector fields
satisfying these conditions is always possible. In this case, the
potential $\theta$ can be expressed as a general smooth function
that depends only on a null coordinate.

The existence of these gravitationally undetectable solutions is
 essentially due to the particular form of the matter action (\ref{action}).
For this reason, we investigate some properties of this matter
action in the second part of this paper. In particular, the
symmetries of the extended Klein-Gordon action (\ref{action}) are
analyzed. It is shown that for the particular value
$\alpha=(D+1)/2$, this action exhibits a conformal invariance for
which the conformal weight of the scalar field $\theta$ is zero. In
contrast with the massless Klein-Gordon action, this conformal
symmetry is achieved without requiring the introduction of a
nonminimal coupling. Finally, we establish a correspondence between
the relativistic action (\ref{action}) and a non-relativistic
isentropic fluid defined in one fewer dimension. The clue of this
correspondence lies in the fact that the non-relativistic space-time
can be viewed as the quotient of a higher-dimensional Lorentz
manifold by the integral curves of a covariantly, lightlike vector
field \cite{Duval:1984cj}. For $\alpha=1/2$, this non-relativistic
fluid is identified with the Chaplygin gas while for $0<\alpha<1/2$,
the model corresponds to the generalized Chaplygin gas. Note that
the Chaplygin cosmology provides an interesting setup to explain
that the expansion of the universe is accelerating
\cite{Kamenshchik:2001cp}, \cite{Bento:2002ps} and
\cite{Banerjee:2005vy}. The Chaplygin gas has also raised a growing
attention because of its connection with the Nambu-Goto action
\cite{Bordemann:1993ep}, its rich symmetrical structure
\cite{Bazeia:1998mg} and its supersymmetric extension
\cite{Jackiw:2000cc}. For a general review on this topics see
\cite{Jackiw:2000mm} and works on supersymmetric fluid models are
reported in \cite{Nyawelo:2001fk}. The features of the Chaplygin gas
can also be understood in a Kaluza-Klein type framework
\cite{Hassaine:1999hn} as well as in its relativistic formulation
\cite{Hassaine:2001ix}. The symmetries of the non-relativistic fluid
are also investigated and it is shown the existence, apart from the
Galileo symmetry, of an additional symmetry whose action is a time
rescaling.

The paper is organized as follows. We first derive solutions of the
equations (\ref{StealthEinsteineqs}) for the energy-momentum tensor
associated to the extended Klein-Gordon action (\ref{action}). In
the remaining sections, we discuss some general features of the
matter action. In particular, we explain the origin of the conformal
invariance of the extended Klein-Gordon action for the particular
value $\alpha=(D+1)/2$. The last part of the paper is devoted to the
correspondence between the relativistic matter action and a
non-relativistic isentropic and polytropic fluid in one fewer
dimension. The symmetries of this non-relativistic fluid are also
analyzed.

\section{Stealth solutions}
We consider the action of $(D+1)-$dimensional gravity coupled to a
scalar field $\theta$ with dynamics described by the extended
Klein-Gordon action (\ref{action}),
\begin{eqnarray}
I=\int d^D\vec{x}\,dt\,\sqrt{-g}\left[\frac{1}{2\kappa}
(R-2\Lambda)-\frac{1}{2}\left(g^{\mu\nu}\,\nabla_{\mu}
\theta\,\nabla_{\nu}\theta\right)^{\alpha}\right],
\end{eqnarray}
where $R$ is the scalar curvature and $\kappa$ is the gravitational
constant. The field equations obtained by varying the metric and the
scalar field read respectively
\begin{subequations}\label{eqs}
\begin{eqnarray}
\label{einseqs} && G_{\mu\nu}+\Lambda
g_{\mu\nu}=\kappa\,T_{\mu\nu},\\
\nonumber\\
\label{we} &&
\nabla_{\mu}\left(\sqrt{-g}\,g^{\mu\nu}\nabla_{\nu}\theta\,
(\nabla_{\sigma}\theta\nabla^{\sigma}\theta)^{\alpha-1}\right)=0,
\end{eqnarray}
\end{subequations}
where the energy-momentum tensor is given by
\begin{eqnarray}
T_{\mu\nu}=2\alpha\nabla_{\mu}\theta\nabla_{\nu}\theta
\left(\nabla_{\sigma}\theta\nabla^{\sigma}\theta\right)^{\alpha-1}-
g_{\mu\nu}\left(\nabla_{\sigma}\theta\nabla^{\sigma}\theta
\right)^{\alpha}. \label{emt}
\end{eqnarray}

We look for a particular solution of the Einstein equations
(\ref{einseqs}) for which both sides of the equations vanish
(\ref{StealthEinsteineqs}). We first show that the equation
$T_{\mu\nu}=0$ implies some restrictions on the scalar field
$\theta$ and on the real parameter $\alpha$. Indeed, independently
of the background metric, the vanishing of the energy-momentum
tensor is possible only if the scalar field satisfies
$\nabla_{\sigma}\theta \nabla^{\sigma}\theta=0$, otherwise the
metric would be proportional to
$$
g_{\mu\nu}\propto\frac{\nabla_{\mu}\theta\nabla_{\nu}\theta}
{\nabla_{\sigma}\theta\nabla^{\sigma}\theta},
$$
and, hence its determinant would vanish. Consequently, the condition
$T_{\mu\nu}=0$ can be consistently achieved only for a scalar field
satisfying $\nabla_{\sigma}\theta \nabla^{\sigma}\theta=0$ and for
$\alpha>1$ because of the form of the energy-momentum tensor
(\ref{emt}). It is interesting to note that under these conditions,
the extended Klein-Gordon equation (\ref{we}) is automatically
satisfied.

In sum, the system described by the particular Einstein equations
(\ref{StealthEinsteineqs}) together with the extended wave equation
(\ref{we}) is equivalent of finding a vacuum metric and a null
vector field that derives from $\theta$,
\begin{eqnarray}
g_{\mu\nu} k^{\mu}k^{\nu}=0,\qquad\qquad
k_{\mu}=\partial_{\mu}\theta. \label{nullv}
\end{eqnarray}
We now show that a static and spherically symmetric spacetime
geometry always admits such vector field (\ref{nullv}). Indeed, let
us consider such geometry which in addition is supposed to solve the
vacuum Einstein equations. The line element is given by
\begin{eqnarray}
ds^2=-f(r)h(r)\,dt^2+\frac{dr^2}{f(r)}+r^2\,d\Sigma_{D-1}^2,
\label{spacetimegeo}
\end{eqnarray}
where $d\Sigma_{D-1}^2$ is the line element of the
$(D-1)-$~dimensional sphere. It is simple to see that the vector
field defined by
\begin{eqnarray}
k^{\mu}=\mp\frac{1}{\sqrt{h(r)}}\delta_r^{\mu}-\frac{1}{f(r)h(r)}\delta_t^{\mu}
\label{nullvector}
\end{eqnarray}
is null and derived from the potential
\begin{eqnarray}
t\mp\int^r\frac{d\tilde{r}}{f(\tilde{r})\sqrt{h(\tilde{r})}}.
\label{pot}
\end{eqnarray}
Hence, any smooth function of this potential (\ref{pot})
\begin{eqnarray}
\theta=F\left(t\mp\int^r\frac{d\tilde{r}}{f(\tilde{r})\sqrt{h(\tilde{r})}}\right),
\label{sol}
\end{eqnarray}
is a solution on the vacuum spacetime metric of the constraint
\begin{eqnarray}
g^{\mu\nu}\nabla_{\mu}\theta\nabla_{\nu}\theta=0. \label{gradient}
\end{eqnarray}
This result is not surprising and can be explained as follows. The
null coordinates $u$ and $v$ associated to the spacetime
(\ref{spacetimegeo}) can be defined as
\begin{subequations}\label{null}
\begin{eqnarray}
\label{u} && u=t-\int^r\frac{d\tilde{r}}{f(\tilde{r})\sqrt{h(\tilde{r})}},\\
\nonumber\\
\label{v} &&
v=t+\int^r\frac{d\tilde{r}}{f(\tilde{r})\sqrt{h(\tilde{r})}},
\end{eqnarray}
\end{subequations}
and, in these coordinates, the metric (\ref{spacetimegeo}) takes the
following form
\begin{eqnarray}
ds^2=-f(r)h(r)\,du\,dv+r^2\,d\Sigma_{D-1}^2. \label{nulspacetimegeo}
\end{eqnarray}
Consequently, any smooth function $F$ of one of the null coordinate,
i. e. $F(u)$ or $F(v)$, will provide a nontrivial solution to the
equation (\ref{gradient}). In the original coordinates
(\ref{spacetimegeo}), this solution becomes precisely (\ref{sol}).

Thus, we have shown that the curvature of vacuum spacetime
(\ref{spacetimegeo}) may not react to the presence of matter source
described by the action (\ref{action}) provided the scalar field is
given by the expression (\ref{sol}). The existence of such
configurations is a direct consequence of the particular form of the
matter action. In what follows, we analyze some properties of this
action.

\section{Properties of the extended Klein-Gordon action}
We first show that the matter action (\ref{action}) can be derived
from the action of a massive complex scalar field with Lagrangian
given by
\begin{eqnarray}
{\cal
L}=\frac{1}{2}g^{\mu\nu}(\partial_{\mu}\Phi)(\partial_{\nu}\Phi^{\star})+
V(\vert\Phi\vert^2), \label{L}
\end{eqnarray}
where $V$ denotes a potential that depends only on the module of the
scalar field. Decomposing the scalar field as
\begin{eqnarray}
\Phi=\frac{f}{m}e^{-im\theta}, \label{decomp}
\end{eqnarray}
where $m$ represents the mass of the complex scalar field, the
Lagrangian density (\ref{L}) becomes
\begin{eqnarray}
{\cal L}=\frac{1}{2m^2}g^{\mu\nu}\partial_{\mu}f\partial_{\nu}f+
\frac{1}{2}f^2\,g^{\mu\nu}\partial_{\mu}\theta\partial_{\nu}\theta+
V(f^2/m^2). \label{cmp}
\end{eqnarray}
As shown below, this expression reduces to the extended Klein-Gordon
Lagrangian (\ref{action}) provided the first term in (\ref{cmp}) is
neglected or eliminated and also for a particular election of the
potential. One possibility to eliminate the first term is to
consider an effective potential that depends also on the derivatives
of the scalar field,
$$
V_{\mbox{\tiny{eff}}}(\vert\Phi\vert,
\partial_{\mu}\vert\Phi\vert)=V(\vert\Phi\vert)-\frac{1}{2}g^{\mu\nu}
(\partial_{\mu}\vert\Phi\vert)(\partial_{\nu}\vert\Phi\vert).
$$
In this case the starting Lagrangian (\ref{L}) with this potential
reduces to
\begin{eqnarray}
{\cal L}=
\frac{1}{2}f^2\,g^{\mu\nu}\partial_{\mu}\theta\partial_{\nu}\theta+
V(f^2/m^2). \label{cmpp}
\end{eqnarray}
One can also reach to this expression (\ref{cmpp}) by neglecting the
first term of (\ref{cmp}) by considering the scale of the
inhomogeneities as corresponding to the spacetime variations of $f$
on scales greater than $m^{-1}$, i. e. $\partial_{\mu}f<<mf$. This
has been considered in the context of cosmology to show that the
generalized Chaplygin gas can be described by a generalized
Born-Infeld Lagrangian \cite{Bento:2002ps}.

The field equations resulting from the variation of the Lagrangian
(\ref{cmpp}) read
\begin{subequations}
\label{fe}
\begin{eqnarray}
\label{fe1}
\frac{1}{2}g^{\mu\nu}(\partial_{\mu}\theta)(\partial_{\nu}\theta)=
-V^{\prime}(f^2),\\
\nonumber\\
 \label{fe2}
\nabla_{\mu}\left(f\sqrt{-g}\,g^{\mu\nu}\nabla_{\nu}\theta\right)=0.
\end{eqnarray}
\end{subequations}
where we have set the mass to unity for simplicity. We consider a
self-interacting potential $V$ given by
\begin{eqnarray}
V(\vert\Phi\vert)=\lambda
\left(\Phi\Phi^{\star}\right)^{\frac{\alpha}{\alpha-1}}
\label{potential}
\end{eqnarray}
where $\lambda$ is a constant. In this case, the action (\ref{cmpp})
becomes
\begin{eqnarray}
S=\int_{M} d^{D}\vec{x}
\,dt\,\sqrt{-g}\left[\frac{1}{2}f^2(\partial_{\mu}\theta)
(\partial^{\mu}\theta)+\lambda
f^{\frac{2\alpha}{\alpha-1}}\right],\label{cmpppp}
\end{eqnarray}
and the function $f$ can be eliminated in favor of the phase
$\theta$ through the equation of motion (\ref{fe1}),
$$
f^2\propto \left(\partial_{\sigma}\theta
\partial_{\sigma}\theta\right)^{\alpha-1}.
$$
Finally, the substitution of this last expression into the
Lagrangian (\ref{cmpppp}) yields the extended Klein-Gordon action
(\ref{action}).

\subsection{Conformal symmetry}
In $(D+1)$ dimensions, the extended Klein-Gordon action
(\ref{action}) possesses the conformal invariance for a value of the
parameter $\alpha$ given by
\begin{eqnarray}
\alpha=\frac{D+1}{2}. \label{calpha}
\end{eqnarray}
Indeed, it is simple to see that the trace of the energy-momentum
tensor (\ref{emt}) vanishes for this value,
$$
g^{\mu\nu}T_{\mu\nu}=\left[2\alpha-(D+1)\right]\left(\partial_{\sigma}\theta
\partial^{\sigma}\theta\right)^{\alpha}.
$$
The implementation of this symmetry on the scalar field $\theta$ is
with a zero conformal weight,
\begin{eqnarray}
g_{\mu\nu}\to \Omega^2\,g_{\mu\nu},\qquad \quad \theta\to\theta.
\label{fields}
\end{eqnarray}
Various comments can be made with respect to this conformal
symmetry. Firstly, in contrast with the standard Klein-Gordon action
this conformal symmetry is achieved without the introduction of the
nonminimal coupling. Moreover, it is interesting to note that the
scalar field action (\ref{action}) remains unchanged against the
transformations (\ref{fields}), i. e.
$$
S_{\frac{D+1}{2}}(\Omega^2\,g_{\mu\nu},\theta)=
S_{\frac{D+1}{2}}(g_{\mu\nu},\theta).
$$
Finally, the origin of this conformal invariance can be explained as
follows. It is well-known that in $(D+1)$ dimension with $D>1$, the
following generalized Klein-Gordon Lagrangian density
\begin{eqnarray}
{\cal L}_{(D,1)}=
\frac{1}{2}g^{\mu\nu}(\partial_{\mu}\Phi)(\partial_{\nu}\Phi^{\star})+
\frac{(D-1)}{8\,D}R\,\Phi\Phi^{\star}\nonumber\\+\lambda\,
\left(\Phi\Phi^{\star}\right)^{\frac{D+1}{D-1}} \label{GKG1}
\end{eqnarray}
is conformally invariant. Here $R$ represents the scalar curvature
of the metric and the last term in (\ref{GKG1}) is the unique
potential that preserves the conformal invariance. For convenience,
we rewrite this action in terms of the module and the phase,
$\Phi=fe^{i\theta}$,
\begin{eqnarray}
{\cal L}_{(D,1)}= \frac{1}{2}\partial_{\mu}f\partial^{\mu}f
                  +\frac{1}{2}f^2\partial_{\mu}\theta\partial^{\mu}\theta
                  +
                  \frac{(D-1)}{8\,D}R f^2\nonumber\\
                  +\lambda\,
                  f^{\frac{2(D+1)}{D-1}},
\label{GKG}
\end{eqnarray}
and the implementation of the conformal symmetry on the dynamical
fields is given by
\begin{eqnarray}
g_{\mu\nu}\to \Omega^2 g_{\mu\nu},\qquad f\to
\Omega^{\frac{1-D}{2}}f,\qquad\theta\to\theta \label{cd}
\end{eqnarray}
A simple but tedious computation shows that the term proportional to
$Rf^2$ is precisely introduced to cancel the variation of the term
$\partial_{\mu}f\partial^{\mu}f$ under the conformal transformation
(\ref{cd}). Consequently, if one drops this kinetic term (as we did
in the previous derivation), the following Lagrangian
\begin{eqnarray}
{\cal L}_{(D,1)} =\frac{1}{2}f^2\,g^{\mu\nu}\partial_{\mu}\theta
\partial_{\nu}\theta+\lambda\, f^{\frac{2(D+1)}{D-1}}
\label{cl}
\end{eqnarray}
also exhibits the conformal invariance. This expression corresponds
precisely to the Lagrangian (\ref{cmpppp}) for the conformal value
of the parameter $\alpha$ given by (\ref{calpha}).

\subsection{Non-relativistic fluid}
We now establish a correspondence between the relativistic action
(\ref{action}) defined in $(D,1)$ dimensions and a non-relativistic
fluid in $(D-1,1)$ dimensions through a Kaluza-Klein type framework.
The clue of this correspondence lies in the fact that the quotient
of a Lorentz $(D,1)-$manifold by the integral curves of a
covariantly constant and lightlike vector field $\xi$ is a
$(D-1,1)-$manifold which carries the geometric structure of
non-relativistic space-time \cite{Duval:1984cj}.

We first illustrate this framework with a simple example. On the
Lorentz manifold we consider a coordinate system given by
$(t,\vec{x},z)$ where $(t,\vec{x})$ are the coordinates on the
non-relativistic space-time and $z$ represents the additional
coordinate. Let $\Phi$ be a complex scalar field satisfying the wave
equation in $(D,1)$ dimensions together with an equivariance
condition
\begin{subequations}
 \label{es}
 \begin{eqnarray}
\Box\Phi=0\\
\xi^{\mu}\partial_{\mu}\Phi=i\Phi.
\end{eqnarray}
\end{subequations}
This system of equations (\ref{es}) has been shown to be strictly
equivalent to the free Schr\"odinger equation in $(D-1,1)$
dimensions on a general Newton-Cartan spacetime \cite{Duval:1984cj}.

In our case, we consider the action defined in (\ref{cmpppp}) which
has been shown to be equivalent to the extended Klein-Gordon action
(\ref{action}). The field equations associated to (\ref{cmpppp})
read
\begin{subequations}
\label{feee}
\begin{eqnarray}
\label{feee1}
\frac{1}{2}g^{\mu\nu}(\partial_{\mu}\theta)(\partial_{\nu}\theta)=
-\frac{\lambda\alpha}{\alpha-1} \rho^{\frac{1}{\alpha-1}},\\
\nonumber\\
 \label{feee2}
\partial_{\mu}\left(\rho\sqrt{-g}\,g^{\mu\nu}\partial_{\nu}\theta\right)=0,
\end{eqnarray}
\end{subequations}
where for convenience we have substituted $f^2$ by $\rho$. In order
to fix the set up of the Kaluza-Klein type framework, we consider
the Minkowski metric written in the light-cone coordinates
\begin{eqnarray}
ds^2=2dtdz+d\vec{x}^2, \label{minklc}
\end{eqnarray}
for which the lightlike vector field $\xi$ can be chosen to be
$\xi^{\mu}\partial_{\mu}=\partial_z$. Since $\rho$ is the module and
$\theta$ the phase of the complex scalar field (\ref{decomp}), the
analogue of the equivariance condition (\ref{es}) is given by
\begin{subequations}
\label{equivariance}
\begin{eqnarray}
\label{equi1} &&\xi^{\mu}\partial_{\mu}\rho=0\Longrightarrow
\rho=\rho(t,\vec{x})\\
\label{dsef} &&\xi^{\mu}\partial_{\mu}\theta=1\Longrightarrow
\theta(t,\vec{x},z)=\Theta(t,\vec{x})+z.
\end{eqnarray}
\end{subequations}
It is easy to see that on the flat background (\ref{minklc}) and for
fields $\rho$ and $\theta$ satisfying the conditions
(\ref{equivariance}), the relativistic field equations (\ref{feee})
project onto the following $(D-1,1)-$dimensional non-relativistic
equations
\begin{subequations}
\label{nrfe}
\begin{eqnarray}
\label{nrfe1}
\partial_t \rho+\vec{\nabla}\cdot(\rho\vec{\nabla}\Theta)=0,\\
\nonumber\\
 \label{nrfe2}
\partial_t\Theta+\frac{1}{2}\vert\vec{\nabla}\Theta\vert^2=-
\frac{\lambda\alpha}{\alpha-1} \rho^{\frac{1}{\alpha-1}}.
\end{eqnarray}
\end{subequations}
These equations turn out to be the non-relativistic equations of an
isentropic, irrotational and polytropic fluid. Indeed, identifying
$\Theta$ as the potential velocity, i. e.
$\vec{v}=\vec{\nabla}\Theta$ and $\rho$ as the density, the first
equation (\ref{nrfe1}) is a continuity relation while the gradient
of the second equation (\ref{nrfe2}) yields an Euler equation of an
isentropic fluid with pressure given by
\begin{eqnarray}
p=\frac{\lambda}{\alpha-1}\rho^{\frac{\alpha}{\alpha-1}}.
\label{poly}
\end{eqnarray}
For a potential strength $\lambda>0$ and for $\alpha=1/2$ (resp. for
$0< \alpha<1/2$), the relation (\ref{poly}) represents the state
equation of the Chaplygin gas (resp. the generalized Chaplygin gas).
Finally, we remark that the field equations (\ref{nrfe}) can be
derived from the following action principle
\begin{eqnarray}
S(\rho,\Theta)=\int d^{D-1}\vec{x}dt\left[\rho
\left(\partial_{t}\Theta+\frac{1}{2}\vert\vec{\nabla}\Theta\vert^2\right)+
\lambda\rho^{\frac{\alpha}{\alpha-1}}\right]. \label{nraction}
\end{eqnarray}

In what follows, we study the symmetries of the non-relativistic
fluid model whose dynamics is described by the equations
(\ref{nrfe}).

\subsubsection{Non-relativistic symmetries}
We analyze the dynamical symmetries of the non-relativistic
isentropic fluid described by the equations (\ref{nrfe}). This model
being non-relativistic possesses the appropriate symmetry, namely
the Galileo symmetry. The action of this symmetry on the coordinates
is given by
\begin{eqnarray}
\left\lbrace
\begin{array}{l}
t\to T=t+\epsilon\nonumber\\
\vec{x}\to \vec{X}={\cal R}\vec{x}+\vec{\delta}-\vec{\beta}t
\end{array}
\right.
\end{eqnarray}
where ${\cal R}\in SO(D-1)$, $\epsilon$, $\vec{\delta}$ and
$\vec{\beta}$ are the parameters associated to the rotations, the
time translations, the space translations and the Galileo boosts
respectively. The implementation of the Galileo symmetry on the
dynamical fields reads
\begin{eqnarray*}
\rho\to \tilde{\rho}(t,\vec{x})=\rho(T,\vec{X}),
\end{eqnarray*}
and
\begin{eqnarray*}
\Theta\to\tilde{\Theta}(t,\vec{x})=\Theta(T,\vec{X})+\vec{\beta}\cdot\vec{x}-
\frac{1}{2}\vert\vec{\beta}\vert^2t.
\end{eqnarray*}
The application of the Noether theorem yields to the following
constants of motion
\begin{subequations}
\label{csts}
\begin{eqnarray}
\label{energy} &H&=\int d^{D-1}\vec{x}\,\,{\cal H}=\int
\left[\frac{1}{2}\rho\vert\vec{\nabla}\Theta\vert^2+
\lambda\rho^{\frac{\alpha}{\alpha-1}}\right]\\
\label{momentum} &\vec{P}&=\int d^{D-1}\vec{x}\,\,{\cal
\vec{P}}=\int d^{D-1}\vec{x}\,\,(\rho\vec{\nabla}\Theta),\\
\label{rotation} &M_{ij}&=\int d^{D-1}\vec{x}\,\,\left(x_i{\cal
P}_j-x_j{\cal P}_i\right),\\
\label{boosts} &\vec{G}&=t\vec{P}-\int
d^{D-1}\vec{x}\,\,(\vec{x}\rho),
\end{eqnarray}
\end{subequations}
which corresponds to the energy $\epsilon$, the momentum
$\vec{\delta}$, the rotations ${\cal R}$ and the Galileo boosts
$\vec{\beta}$ respectively. The equations are also invariant under a
shift of the velocity potential by constant,$\Theta\to
\Theta+\mbox{constant}$, and the associated conserved charge is
given by
\begin{eqnarray}
N=\int d^{D-1}\vec{x}\,\,\rho. \label{N}
\end{eqnarray}
The corresponding Lie algebra generated by $H, \vec{P}, M_{ij},
\vec{G}$ and $N$ is the Galileo algebra with a central extension
given by (\ref{N}) and corresponds to the Galileo $2-$cocycle.

$\bullet$ For a generic value of the parameter $\alpha$, there
exists an additional symmetry which does not belong to the Galileo
group. This symmetry acts on the coordinates by rescaling only the
time
\begin{eqnarray}
&&t\to T=e^{\omega}\,t,\nonumber \\
&&\vec{x}\to \vec{X}=\vec{x}, \label{timedil}
\end{eqnarray}
while its action on the dynamical fields is given by
\begin{eqnarray}
&&\rho\to \tilde{\rho}(t,\vec{x})=e^{2(\alpha-1)\omega}\,\rho(T,\vec{X}),\nonumber\\
&&\Theta\to\tilde{\Theta}(t,\vec{x})=e^{\omega}\,\Theta(T,\vec{X}),
\label{chge}
\end{eqnarray}
or infinitesimally
\begin{eqnarray}
&&\delta\rho=2(\alpha-1)\rho(t,\vec{x})+t\partial_t\rho(t,\vec{x}),\nonumber\\
&&\delta{\Theta}=\Theta(t,\vec{x})+t\partial_t\Theta(t,\vec{x}).
\label{chgee}
\end{eqnarray}

 The associated conserved quantity reads
\begin{eqnarray}
B_{\alpha}=tH-\frac{(3-2\alpha)}{2\alpha+1}\int
d^{D-1}\vec{x}\,\,\left(\rho\,\Theta\right)-\nonumber\\
\frac{(2\alpha-1)}{2\alpha+1}\int
d^{D-1}\vec{x}\,\,\left(\vec{x}\cdot\vec{\nabla}\Theta\right)\rho.
\label{brasil}
\end{eqnarray}
This expression is clearly not defined for $\alpha=-1/2$ in spite of
the fact that this value is not singular at the level of the
transformations (\ref{timedil}) and (\ref{chge}). In fact, a careful
application of the Noether procedure shows that for $\alpha=-1/2$
the associated conserved quantity does not involve the energy
density and instead is given by
\begin{eqnarray}
B_{-1/2}=\int d^{D-1}\vec{x}\,\,\left[\rho\Theta-\frac{1}{2}
\left(\vec{x}\cdot\vec{\nabla}\Theta\right)\rho\right].
\label{brasil-1/2}
\end{eqnarray}
In the case of the Chaplygin gas, $\alpha=1/2$, the last piece of
the expression (\ref{brasil}) vanishes and the Noether charge
corresponds to the one derived in \cite{Bazeia:1998mg}.
Interestingly, for $\alpha\not=1/2$, the two conserved charges
(\ref{brasil}) and (\ref{brasil-1/2}) involve a piece proportional
to the space coordinate $\vec{x}$. This fact is intriguing since the
transformations (\ref{timedil}) only affect the time and not the
space coordinate. This can be explained by the fact that, under the
infinitesimal changes of the dynamical fields (\ref{chgee}), the
variation of the action (\ref{nraction}) becomes
\begin{eqnarray}
\label{var} \Delta S &=&
S(\rho+\delta\rho,\Theta+\delta\Theta)-S(\rho,\Theta)\\
&=&(2\alpha-1)S(\rho,\Theta)+\int\,\,\partial_{t}\left[\rho\partial_{t}(t\Theta)-
\rho\Theta+t{\cal H}\right]\nonumber,
\end{eqnarray}
where ${\cal H}$ represents the density energy (\ref{energy}). For
$\alpha=1/2$, this variation (\ref{var}) reduces to a surface term
and, hence a direct application of the Noether theorem yields to the
conserved charge (\ref{brasil}). For $\alpha\not=1/2$, the reason
for which the transformation (\ref{timedil}) still acts as a
symmetry is due to the fact that the original action can also be
written as a surface term. Indeed, using the equations of motion
(\ref{nrfe}), the action (\ref{nraction}) can be expressed on-shell
as
\begin{eqnarray}
S(\rho,\Theta)=\frac{1}{2\alpha+1}\int
d^{D-1}\vec{x}\,dt\,\,\,\partial_{t}\left(2\rho\,\Theta-\rho\,
\vec{x}\cdot\vec{\nabla}\Theta \right) \label{ds}
\end{eqnarray}
As a consequence, for $\alpha\not=1/2$, the variation $\Delta S$ of
the action (\ref{var}) under the time dilatation transformation
(\ref{timedil}) is also a surface term. This proves that these
transformations act as a symmetry for the action and also the reason
for which the conserved quantities (\ref{brasil}) and
(\ref{brasil-1/2}) involve the space coordinate $\vec{x}$.

For a generic value of the parameter $\alpha$, the group structure
of the symmetries is given by the semidirect sum of the Galileo
group with central extension ${\cal G}$ with the generator
associated to this extra symmetry (\ref{brasil}),
\begin{eqnarray}
G={\cal G}\ltimes {\cal B}. \label{lo}
\end{eqnarray}

There exist two particular values of the parameter $\alpha$ for
which the symmetry group can be larger than the one discussed before
(\ref{lo}).

$\bullet$ For the value $\alpha=1/2$, which corresponds to the
Chaplygin gas, it has been shown that, apart from the time dilation
(\ref{timedil}), there exists an extra symmetry whose action is
field dependent, i. e.
\begin{eqnarray}
\begin{array}{c}
t\to
T=t+\frac{1}{2}\vec{\omega}\cdot(\vec{x}+\vec{X})\\
\\
\vec{x}\to \vec{X}=\vec{x}+\vec{\omega}\Theta(T,\vec{X})
\end{array}
\end{eqnarray}
and
\begin{eqnarray}
\begin{array}{c}
\rho\to \tilde{\rho}(t,\vec{x})=\rho(T,\vec{X})
\displaystyle{\frac{1}{\vert J\vert}}\\
\\
\Theta\to \tilde{\Theta}(t,\vec{x})=\Theta(T,\vec{X})
\end{array}
\end{eqnarray}
where $\vert J\vert$ is the Jacobian of the transformation,
$J=\mbox{det}\left(\partial X^{\mu}/\partial x^{\sigma}\right)$,
\cite{Bazeia:1998mg}. The Noether conserved quantity is given by
\begin{eqnarray}
\vec{D}=\int d^{D-1}\vec{x}\,dt\,\,\,\left(\vec{x}\,{\cal
H}-\Theta\,{\cal \vec{P}}\right), \label{antiboosts}
\end{eqnarray}
and the group structure generated by the quantities (\ref{csts}),
(\ref{brasil}) and (\ref{antiboosts}) is the Poincar\'e group in one
higher dimension, namely in $(D,1)$ dimensions. The existence of
this symmetry is due to the fact that for $\alpha=1/2$, the action
expressed in terms of $\Theta$ can be written as a square root,
$$
S=\int
d^{D-1}\vec{x}\,dt\,\,\,\sqrt{\partial_t\Theta+\frac{1}{2}\vert
\vec{\nabla}\Theta\vert^2},
$$
and this later can be seen to descend from a Nambu-Goto action in
one higher dimension in the light cone parametrization
\cite{Bordemann:1993ep}. This explains the arising of the Poincar\'e
symmetry only for $\alpha=1/2$ since for the other values of
$\alpha$, the action is not a square root and hence it is not longer
parametrization invariant.

$\bullet$ In $(D-1,1)$ dimensions and for a value of $\alpha$ given
by $(D+1)/2$, the polytropic fluid exhibits a Schr\"odinger symmetry
for which the group structure is the semidirect sum of the static
Galileo group with $SL(2,I\!\! R)$, (see
\cite{O'Raifeartaigh:2000mp}, \cite{Henkel:2003pu} and for an
extension to discontinuous flows see \cite{Jahn:2004di}). The
arising of this symmetry is a consequence of the conformal symmetry
of the relativistic model in one higher dimension.

\section{Discussion}
Here, we have considered the Einstein equations with a matter source
scalar field. The dynamics of the scalar field is described by an
extended Klein-Gordon action that depends on a real parameter
$\alpha$. We have restricted ourselves to a particular class of
solutions for which both sides of the Einstein equations vanish. In
this case, we have shown that for any static, spherically symmetric
vacuum solutions of the Einstein equations, a nontrivial scalar
field with zero energy-momentum on-shell can be derived. This means
that this matter source in spite of being coupled to gravity does
not affect the curvature of the spacetime. An interesting work will
consist to see whether scalar field with a non-zero energy momentum
tensor can be coupled to black hole geometry. The existence of these
undetectable solutions is essentially due to the form of the matter
action. For this reason, some properties of this action have been
analyzed. In particular, we have shown that the extended
Klein-Gordon action possesses a conformal invariance for a
particular value of the exponent $\alpha$. This invariance does not
require the introduction of the nonminimal coupling as it is the
case for the standard Klein-Gordon action. We have also established
a correspondence between this extended Klein-Gordon dynamics and a
non-relativistic isentropic fluid in one fewer dimension. This gas
can be identified with the (generalized) Chaplygin gas for specific
values of the parameter $\alpha$. The non-relativistic model
corresponds to an irrotational, isentropic and polytropic fluid.
This fluid admits, apart from the Galileo symmetry, an additional
symmetry. The action of this extra symmetry on the coordinate
consists of a rescaling of the time only. An interesting work will
consist of finding solutions of this non-relativistic fluid and to
make use of this extra symmetry to generate non trivial solutions.

\bigskip

{\bf Acknowledgments.-} We thank E. Ay\'on-Beato, A. Gomberoff, C.
Mart\'{\i}nez, R. Portugues, R. Troncoso and J. Zanelli for useful
discussions. This work is partially supported by grants 1051084 and
1060831 from FONDECYT. Institutional support to the Centro de
Estudios Cient\'{\i}ficos (CECS) from Empresas CMPC is gratefully
acknowledged. CECS is a Millennium Science Institute and is funded
in part by grants from Fundaci\'{o}n Andes and the Tinker
Foundation.



\begin{thebibliography}{99}
\bibitem{S3d} M.~Natsuume, T.~Okamura, and M.~Sato, Phys.\ Rev.\ D \textbf{61}, 104005
(2000); E.~Ay\'{o}n-Beato, A.~Garc\'{\i}a, A.~Mac\'{\i}as, and
J.M.~P\'{e}rez-S\'{a}nchez, Phys.\ Lett.\ B \textbf{495}, 164
(2000); M.~Henneaux, C.~Mart\'{\i}nez, R.~Troncoso, and J.~Zanelli,
Phys.\ Rev.\ D \textbf{65}, 104007 (2002); J.~Gegenberg,
C.~Mart\'{\i}nez, and R.~Troncoso, Phys.\ Rev.\ D \textbf{67},
084007 (2003); E.~Ay\'{o}n--Beato, C.~Mart\'{i}nez, and J.~Zanelli,
arXiv:hep-th/0403228.

\bibitem{Ayon-Beato:2005tu} E.~Ayon-Beato, C.~Martinez, R.~Troncoso and J.~Zanelli, Phys.\ Rev.\ D {\bf 71},
104037 (2005).


\bibitem{Duval:1984cj} C.~Duval, G.~Burdet, H.~P.~Kunzle and M.~Perrin, Phys.\ Rev.\ D {\bf 31}, 1841
(1985); C.~Duval, G.~W.~Gibbons and P.~Horvathy, Phys.\ Rev.\ D {\bf
43}, 3907 (1991).


\bibitem{Kamenshchik:2001cp} A.~Y.~Kamenshchik, U.~Moschella and V.~Pasquier, Phys.\ Lett.\ B {\bf 511}, 265
(2001).

\bibitem{Bento:2002ps}
M.~C.~Bento, O.~Bertolami and A.~A.~Sen, Phys.\ Rev.\ D {\bf 66},
043507 (2002).




\bibitem{Banerjee:2005vy} R.~Banerjee and S.~Ghosh, {\it New approach to Chaplygin gas cosmology}, arXiv:gr-qc/0508021.




\bibitem{Bordemann:1993ep} M.~Bordemann and J.~Hoppe, Phys.\ Lett.\ B {\bf 317}, 315
(1993).

\bibitem{Bazeia:1998mg} D.~Bazeia and R.~Jackiw, Annals Phys.\  {\bf 270}, 246
(1998); R.~Jackiw and A.~P.~Polychronakos, Commun.\ Math.\ Phys.\
{\bf 207}, 107 (1999).

\bibitem{Jackiw:2000cc} R.~Jackiw and A.~P.~Polychronakos, Phys.\ Rev.\ D {\bf 62}, 085019
(2000); Y.~Bergner and R.~Jackiw, Phys.\ Lett.\ A {\bf 284}, 146
(2001).

\bibitem{Jackiw:2000mm} R.~Jackiw, {\it A particle field theorist's lectures on supersymmetric,
non-Abelian  fluid mechanics and d-branes}, arXiv:physics/0010042;
R.~Jackiw, V.~P.~Nair, S.~Y.~Pi and A.~P.~Polychronakos, J.\ Phys.\
A {\bf 37}, R327 (2004).

\bibitem{Nyawelo:2001fk} T.~S.~Nyawelo, J.~W.~Van Holten and S.~Groot Nibbelink, Phys.\ Rev.\ D {\bf 64},
021701 (2001);  A.~Das and Z.~Popowicz, Phys.\ Lett.\ A {\bf 296},
15 (2002); T.~S.~Nyawelo, Nucl.\ Phys.\ B {\bf 672}, 87 (2003).


\bibitem{Hassaine:1999hn} M.~Hassaine and P.A.Horvathy, Annals Phys.\  {\bf 282}, 218
(2000);  M.~Hassaine and P.~A.~Horvathy, Lett.\ Math.\ Phys.\ {\bf
57}, 33 (2001); M.~Hassaine, Phys.\ Lett.\ A {\bf 290}, 157 (2001).

\bibitem{Hassaine:2001ix} M.~Hassaine and P.~A.~Horvathy, Lett.\ Math.\ Phys.\  {\bf 57}, 33
(2001).


\bibitem{O'Raifeartaigh:2000mp} L.~O'Raifeartaigh and V.~V.~Sreedhar, Annals Phys.\  {\bf 293}, 215
(2001);  M.~Hassaine and P.~A.~Horvathy, Phys.\ Lett.\ A {\bf 279},
215 (2001).


\bibitem{Henkel:2003pu} M.~Henkel and J.~Unterberger, Nucl.\ Phys.\ B {\bf 660}, 407
(2003).


\bibitem{Jahn:2004di} O.~Jahn, V.~V.~Sreedhar and A.~Virmani, Annals Phys.\  {\bf 316}, 30
(2005).


\end{thebibliography}
\end{document}